\documentclass[12pt]{iopart}
\usepackage[dvips]{graphicx}
\usepackage{xfrac}
\usepackage{amssymb}
\newcommand{\be}{\begin{equation}}
\newcommand{\ee}{\end{equation}}
\newcommand{\bea}{\begin{align}}
\newcommand{\eea}{\end{align}}
\def\ket#1{|#1\rangle}
\def\bra#1{\langle#1|}
\def\V{V}

\def\Vs{\mathbb{C}^{2s+1}}
\def\Vi{V_i}
\def\Us{U_s}
\def\UV{U_V}
\def\refeq#1{(\ref{#1})}
\def\J{j}
\def\tj#1#2#3#4#5#6{\left(\hbox{\begin{tabular}{ccc}$#1$&$#2$&$#3$\\$#4$&$#5$&$#6$\end{tabular}}\right)}
\def\sj#1#2#3#4#5#6{\left\{\hbox{\begin{tabular}{ccc}$#1$&$#2$&$#3$\\$#4$&$#5$&$#6$\end{tabular}}\right\}}
\def\Jl{j_l}
\def\Jr{j_r}
\def\il{i_l}
\def\ir{i_r}
\def\til{\tilde i_l}
\def\tir{\tilde i_r}
\def\s#1#2{\sfrac{#1}{#2}}

\def\12{\frac{1}{2}}
\def\32{\frac{3}{2}}

\begin{document}

\title{Diagrammatics for $SU(2)$ invariant matrix product states}

\author{Andreas, Fledderjohann, Andreas Kl\"umper, Karl-Heinz M\"utter}
\address{Fachbereich C\ Physik, Bergische Universit\"at  Wuppertal,
42097 Wuppertal, Germany}


\begin{abstract}
We report on a systematic implementation of $su(2)$ invariance for matrix
product states (MPS) with concrete computations cast in a diagrammatic
language.  As an application we present a variational MPS study of a spin-1/2
quantum chain. For efficient computations we make systematic
use of the $su(2)$ symmetry at all steps of the calculations: (i) the matrix
space is set up as a direct sum of irreducible representations, (ii) the local
matrices with state-valued entries are set up as superposition of $su(2)$
singlet operators, (iii) products of operators are evaluated algebraically by
making use of identities for $3j$ and $6j$ symbols. The remaining numerical
computations like the diagonalization of the associated transfer matrix and
the minimization of the energy expectation value are done in spaces free of
symmetry degeneracies. The energy expectation value is a strict upper bound of
the true ground-state energy and yields definite conclusions about the
accuracy of DMRG results reported in the literature. Furthermore, we present
explicit results with accuracy better than $10^{-4}$ for nearest- and
next-nearest neighbour spin correlators and for general dimer-dimer
correlators in the thermodynamical limit of the spin-$\12$ Heisenberg chain
with frustration.
\end{abstract}

\maketitle

\section{Introduction}

We use a systematic implementation of $su(2)$ invariance for
matrix-product states (MPS), which parallels \cite{Dukelsky98EPL,Dukelsky98JPA},
and perform a variational $su(2)$-MPS study for the frustrated
antiferromagnetic spin-$\12$ Heisenberg chain.

The class of MPS is of particular importance for the study of quantum spin
chains. The MPS appear in different, but closely related ways. Historically,
products of matrices with entries from a local Hilbert space first appeared as
exact ground-states (matrix-product ground-states, MPG) for Hamiltonians with
special local ground-state structure, see \cite{AKLT87,AKLT88} and
developments \cite{Fannes89}-\cite{KolezhukMY97}. Second, in the literature on
integrable systems, vertex-operators \cite{JMMN92}-\cite{BoosGKS06} were
introduced for a transfer-matrix like construction of ground-states of lattice
systems.  Third, the density-matrix renormalization group (DMRG)
\cite{White92}-\cite{Schollwoeck05} algorithms were shown
\cite{OstlundR95,RommerO97} to result in states of MPS type.  The MPG,
vertex-operators and DMRG are all realizations of MPS. Consequently, this
important class of states attracted strong interest in the quantum computation
community \cite{VerstraeteCLRW05}-\cite{SchuchWVC08}.

In applications of MPS, details may differ strongly. For instance, in MPG and
DMRG realizations the matrix index space is finite dimensional, whereas in the
vertex-operator case this space is mostly infinite dimensional. In MPG, the
MPS are used as an ansatz for an ``exact state'' for which a ``parental
Hamiltonian'' is to be found in subsequent investigations. In DMRG, the MPS
appear as variational states for the Hamiltonian. In case of the
vertex-operator, the full ground-state structure is captured at the expense of
infinite-dimensional matrices.

Our investigation is motivated by the need for a computationally most
efficient scheme for general $su(2)$ singlet states of MPS type. The local
implementation of Lie group invariance has been studied in for instance
\cite{Sanz09} where $su(2)$ invariant MPG and parental Hamiltonians were
investigated, and particularly in the early work
\cite{Dukelsky98EPL,Dukelsky98JPA} where also the variational analysis of
$su(2)$ invariant MPS for quantum chains and ladders was introduced. The paper
\cite{Dukelsky98EPL} is well-known for the insight that the finite-system DMRG
leads to quantum states in MPS form, over which it variationally optimizes. To
our knowledge the computational scheme of $su(2)$ invariant MPS with arbitrary
matrix space did not attract the attention the papers
\cite{Dukelsky98EPL,Dukelsky98JPA} deserve. Non-Abelian symmetries were
implemented in DMRG studies in for instance \cite{McCulloch02,Dukelsky04}.


Here we use a variational computation scheme very similar to
\cite{Dukelsky98EPL,Dukelsky98JPA} based on $su(2)$ invariant MPS. However, we
reduce the necessary constructions to a minimum without taking any reference
to DMRG algorithms. Also, we are going to use a diagrammatic representation of
some of the key objects and relations occurring in the process of the
evaluation of the norm and the energy expectation value of the MPS. We hope, this
approach will make the subject as accessible as possible. Our main
application will be the study of ground-state energies, spin-spin and
dimer-dimer correlations of the (frustrated) antiferromagnetic spin-1/2
Heisenberg chain with nearest- and next-nearest neighbour interactions. For
not too strong frustration, this system shows critical behaviour. Still, the
variational $su(2)$ invariant MPS give excellent results even for the
correlation functions. Probably, and in contrast to the expectation expressed
in \cite{Dukelsky98JPA}, the approach is applicable even to odd-legged
spin-1/2 ladders that are not finitely correlated.

Obviously, and following scientific lore, it is important to use all available
symmetries to find invariant blocks of the transfer matrix as low dimensional
as possible in order to reduce the computational work involved with the
diagonalization procedure. Even more important than the economical treatment
of the transfer matrix is the efficient, nonredundant parameterization of the
local building elements,~i.e. the matrices with entries from the local Hilbert
space. It is essential to parameterize these objects with as few parameters as
possible to reduce the computational time of the minimization of the energy
expectation value.

The paper is organized as follows. In Sect.~\ref{su2inv} we present a fairly
general derivation of equilibrium states in the form of MPS and shortly
summarize the tensor calculus of MPS with emphasis on realizations of
symmetries. In Sect.~\ref{3j6j} we introduce the $su(2)$ invariant local
objects based on Wigner's $3j$ symbols. Here we also introduce the transfer
matrices and evaluate products of operators by making use of identities
involving $3j$ and $6j$ symbols. In Sect.~\ref{Results} we present explicit
results from numerical evaluations of the basic formulas derived in the
previous section. The results are compared with DMRG data of the literature
\cite{Chitra95}
for the frustrated spin-$\12$ Heisenberg chain and conclusions about the
accuracy of the methods are drawn.

\section{Derivation of matrix product states and realization of $SU(2)$ invariance}\label{su2inv}

We are going to study quantum spin systems with local interactions. It is
well-known that quantum system in $d$ spatial dimensions can be mapped to
classical systems in $d+1$ dimensions. In this way, the ground-state
properties of a quantum chain in the thermodynamical limit are encoded by a
classical system on an unrestricted 2-dimensional square lattice. Often, for
numerical purposes, the quantum chain is mapped onto a kind of Ising model on
a square lattice with chequerboard structure, for analytical purposes the
mapping of quantum chains onto vertex models on periodic square lattices is
more convenient. (The associated classical vertex model has nearest-neighbour
couplings even for quantum spin-chains with interactions ranging farther
than nearest-neighbours.) Here, the reasoning is based on equilibrium states,
but obviously the derivation is more general and covers all steady state
systems with local interactions.

After mapping the quantum chain onto a vertex model, the
correlation functions of the classical model on the full plane yield the
correlations of the quantum chain, as is well-known. It is less well-known,
but equally easy to understand that the partition function of a half-plane with
arbitrary, but fixed boundary spins yields the coefficients of the ground-state
of the quantum chain with respect to the standard basis. The evolution
operator associated with a column of the full plane is known as the
transfer matrix of the model. The corresponding objects of the half-plane are
known as (lattice) vertex operators. 
\begin{figure}
\begin{center}
\includegraphics[width=0.9\linewidth]{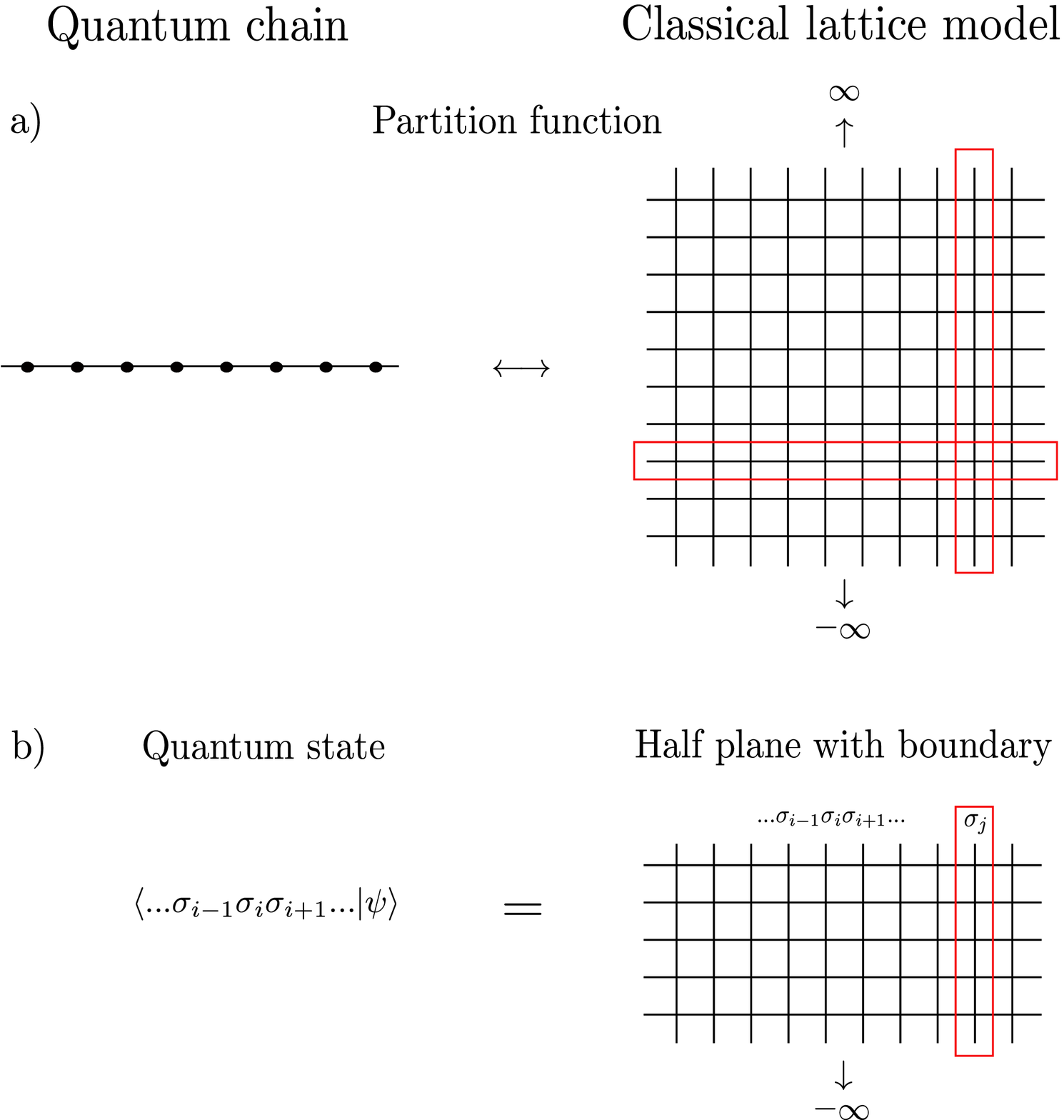}
\end{center}
\caption{a) Illustration of the relation of quantum spin chains and classical
  models on square lattices. The rows and columns of the square lattice define
  the action of the row-to-row and the column-to-column transfer matrix. b)
  The coordinates $\langle
  ...\sigma_{i-1}\sigma_{i}\sigma_{i+1}... |\psi\rangle$ of the ground-state
  of the quantum system correspond to the partition functions of the classical
  model on the half-plane with fixed boundary spins 
  $  ...\sigma_{i-1}\sigma_{i}\sigma_{i+1}...$. The column of the half-plane
  with fixed boundary defines the vertex operator.}
\label{quantumclassical}
\end{figure}
With view to Fig.~\ref{quantumclassical} these objects carry
spin variables on the left, on the right and one spin variable on the
top. When considering these objects as matrices where the spin variables on
the left play the role of the row index, and the spin variables on right play
the role of the column index, vertex operators are matrices with (local) spin
state valued entries.  The goal of the DMRG procedure may be understood as the
computation of the optimal truncation of the infinite dimensional vertex
operator to a finite dimensional matrix space.  In this section 
we present the algebraic
background for the construction of general $su(2)$ invariant MPS.

We first summarize the basic algebraic constructions needed for our investigation
of many-body quantum spin systems. We place particular emphasis on the
compatibility with symmetry groups notably Lie groups. Eventually we will be
interested in the Lie group $SU(2)$ which is the reason for being specific
from the beginning.

We consider the class of matrix-product states
\be
\ket{\psi}=\Tr (g_1\cdot g_2\cdot ...\cdot g_L)\label{MPG}
\ee
where $g_i$ is a square matrix with some auxiliary (index) space $\V$ and
entries from a local quantum space $\Vi$ which we take as the $i$th copy of 
a $su(2)$ spin-$s$ space $\mathbb{C}^{2s+1}$.

$SU(2)$-invariance of $\ket{\psi}$ is guaranteed if for the representation
$\Us$ of $SU(2)$ in $\mathbb{C}^{2s+1}$ there is a representation $\UV$ in $V$
such that $\Us$ applied to any element of the matrix $g_i$, denoted by $\Us
g_i$, yields a matrix identical to $\UV^{-1} \cdot g_i \cdot \UV$ where dots
refer to matrix multiplication. Obviously we obtain with \refeq{MPG}
\begin{eqnarray}
\nonumber
\Us\otimes\Us\otimes...\otimes\Us\ket{\psi}&=\Tr (\Us g_1\cdot \Us g_2\cdot
...\cdot \Us g_L)\\\nonumber
&=\Tr (\UV^{-1} \cdot g_1\cdot \UV \cdot \UV^{-1} \cdot g_2\cdot \UV\cdot
...\cdot \UV^{-1} \cdot g_L\cdot \UV)\\
=\ket{\psi}\label{MPGsym}
\end{eqnarray}

The local condition for $SU(2)$-invariance can be written as
\be
\UV\otimes\Us\ g\ \UV^{-1} = g,\label{gsym1}
\ee
meaning that the object $g$ may be regarded as a tensor of the space
$V\otimes\Vs\otimes V^*$ where $V^*$ is the dual space to $V$. (Note that the
product $A\cdot B$ of two linear maps $A$ and $B$ of the space $V$ corresponds to the
tensor product followed by a contraction of $A$ and $B$ viewed as tensors in
$V\otimes V^*$.)

For the purpose of imposing discrete lattice symmetries like parity,~i.e. 
invariance with respect to reflections, we adopt a different point of
view. Let us consider tensors $G$ from $V\otimes\Vs\otimes V$. As a local
condition for $SU(2)$-invariance we demand
\be
\UV\otimes\Us\otimes\UV\ G = G,\label{Gsym1}
\ee
and as a local condition for parity invariance we demand -- as a sufficient
condition -- that $G$ be symmetric with respect to exchange of ``the first and
the third index'' when written in a canonical basis.

The relation between $g$ and $G$ is realized by a $SU(2)$-invariant tensor $S$
from $V\otimes V$ (and by the invariant dual tensor $S^*$ from $V^*\otimes V^*$).
Concrete candidates for $S$ (and $S^*$) will be given shortly. 

The tensor $S$ can equivalently be understood as a linear map from $V^*$ to $V$
as the multiplication of an arbitrary element $\tilde v$ of $V^*$ with $S$
yields an object in $V^*\otimes V\otimes V$, the subsequent contraction over
the first and second space yields an element $v$ of $V$. Denoting $S(\tilde
v):=v$ we establish $S$ as a map $V^*\to V$. The $SU(2)$-invariance of $S$ as a
tensor in $V\otimes V$ is written as $\UV\otimes\UV\ S=S$ from which we find
\be
S(\tilde v \UV^{-1})=\UV v=\UV S(\tilde v).\label{invarS}
\ee

The object $G$ is obtained by action of $S$ on the third space of $g$
\be
G={\rm id}\otimes{\rm id}\otimes S\ g
\ee
which takes \refeq{gsym1} into \refeq{Gsym1} thanks to \refeq{invarS}.

Conversely, we establish $S^*$ as a linear map $V\to V^*$ with
$SU(2)$-invariance $S^*(\UV v)=S^*(v)\UV^{-1}$ and for the concrete
realizations of $S$ and $S^*$ we find $S\cdot S^*=(-1)^{2\J}$ and $S^*\cdot
S=(-1)^{2\J}$ in spin-$\J$ subspaces. Hence, $S$ and $S^*$ are invertible and 
$g={\rm id}\otimes{\rm id}\otimes S^{-1}\ G$.

Finally, we have to give explicit constructions for $S$ (and $S^*$) as
$SU(2)$-invariant states in $V\otimes V$. This is the only place where we
make explicit use of the fact that our Lie symmetry group is $SU(2)$. We take
the space $V$ as a direct sum of some irreducible spin-$\J$ representations
where $\J=0, \12, 1, \32, 2,...$ . Each $\J$ may appear an arbitrary number of
times, in which
case we label the different orthogonal mupltiplets by an integer $i$. The
space $V$ is spanned by orthogonal states $\ket{(\J,i),m}$ where the magnetic
quantum number $m$ varies from $-\J$ to $+\J$ in integer steps. $SU(2)$ singlet
states in $V\otimes V$ and $V^*\otimes V^*$ are given by
\begin{eqnarray}
\nonumber
S:=\sum_{\J,i}\sum_{m=-\J}^\J(-1)^{\J-m}\ket{(\J,i),m}\otimes\ket{(\J,i),-m},\\
S^*:=\sum_{\J,i}\sum_{m=-\J}^\J(-1)^{\J-m}\bra{(\J,i),m}\otimes\bra{(\J,i),-m}.
\end{eqnarray}
where $S^*$ is related to $S$ by replacing the ket-states by the dual bra-states.

Applying our above formulated definitions we find for $\tilde v=\bra{(\J,i),m}$
that $S(\tilde v)=(-1)^{\J-m}\ket{(\J,i),-m}$. Conversely, for $v=\ket{(\J,i),m}$
we have $S^*(v)=(-1)^{\J-m}\bra{(\J,i),-m}$. Hence, the successive action of $S$ and
$S^*$ yields $(-1)^{2\J}\, {\rm id}$.

\section{Basic representation theoretical settings: $3j$ and $6j$ 
symbols}\label{3j6j}

Having spelled out the fundamental objects appearing as factors in
$SU(2)$-invariant matrix-product states, the concrete calculations are
straightforward. We want to use $SU(2)$ singlets $G$ in $V\otimes\Vs\otimes
V$. Having already allowed for reducible representations in $V$, we like to
stress that for our applications we {\em must} deal with $V$ as direct sum of
more than one irreducible representation. This is so as for the most
interesting case of $s=\12$ no singlet $G$ exists if $V$ is identical to just
one spin-$\J$ multiplet. There is no half-integer spin in the tensor product
decomposition of two spin-$\J$'s!

\begin{figure}
\begin{center}
\includegraphics[width=0.8\linewidth]{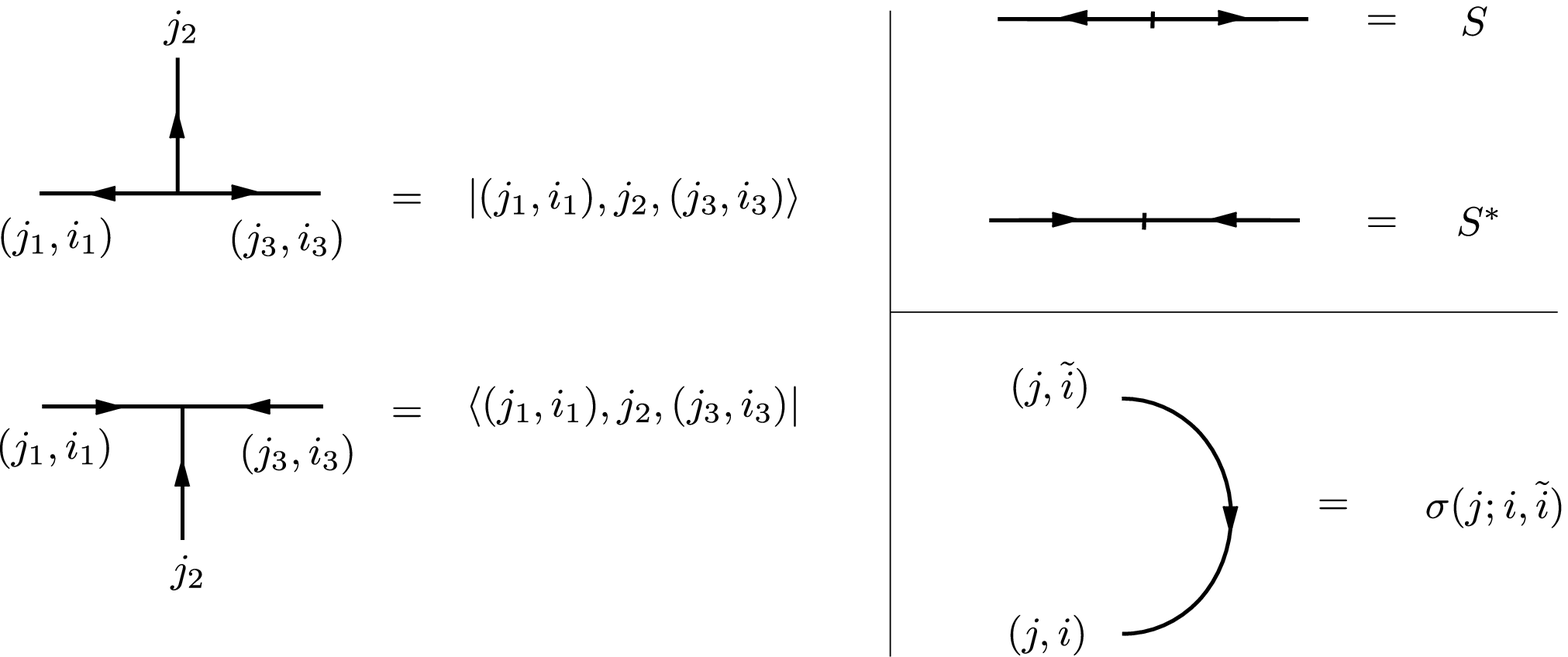}
\end{center}
\caption{Introduction of the graphical notation for singlets appearing in
product spaces. Three leg (vertex)
symbols for $\ket{(\J_1,i_1),\J_2,(\J_3,i_3)}$ and
$\bra{(\J_1,i_1),\J_2,(\J_3,i_3)}$ appearing in 
products of three irreducible representations. Two leg (edge) symbols for the
singlets $S$ resp. $S^*$ in $V\times V$ resp. $V^*\times V^*$, and arcs for 
singlets in $V\times V^*$ for the same spin $j$.}
\label{graphsarrows}
\end{figure}

\begin{figure}
\begin{center}
\includegraphics[width=0.6\linewidth]{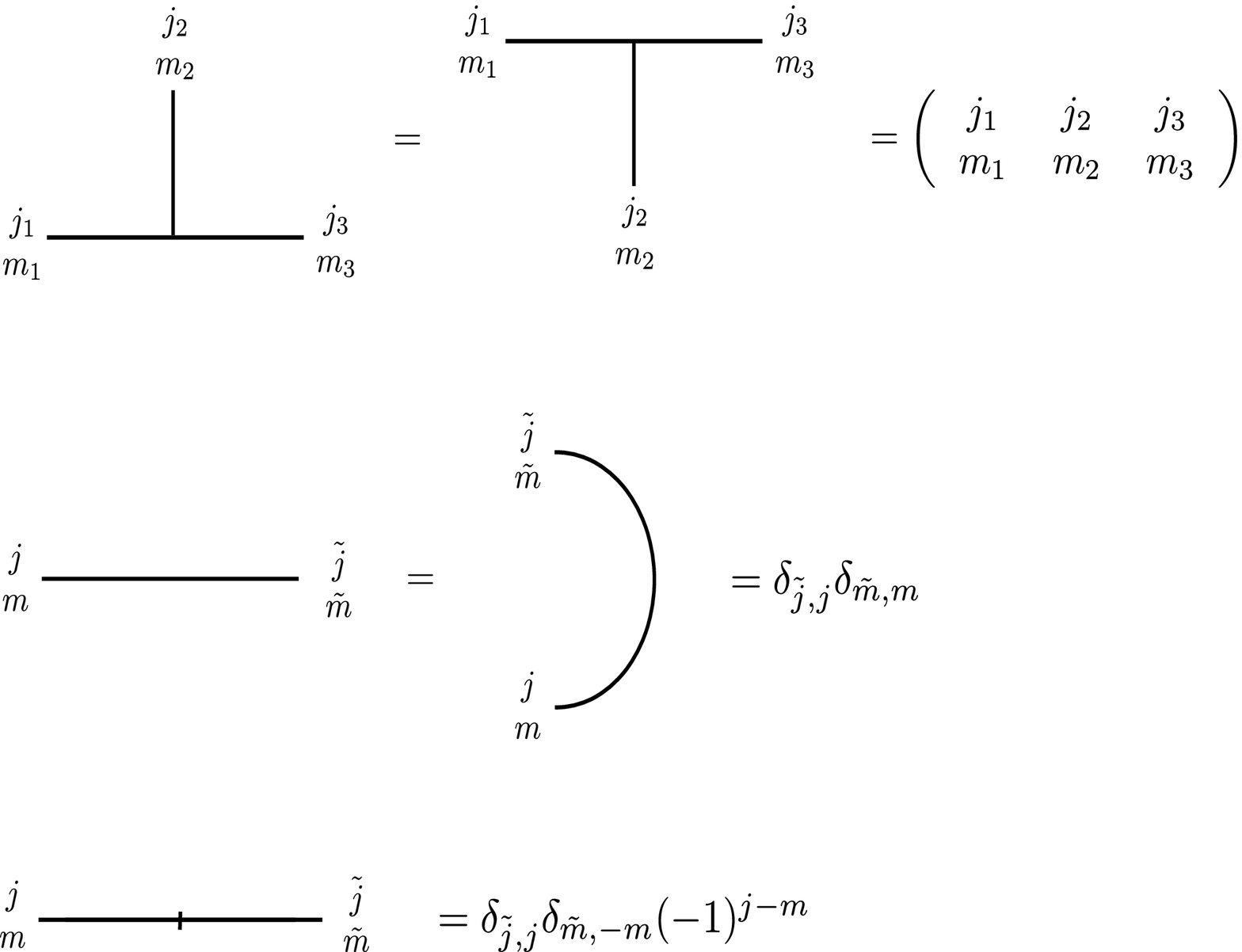}
\end{center}
\caption{Graphical notation for key arithmetic objects like Kronecker symbol
  and $3j$ symbol.}
\label{graphsnoarrows}
\end{figure}

Let us consider in $V\otimes\Vs\otimes V$ any spin multiplet $(\J_1,i_1)$ from
the first factor space, the (only) spin multiplet $\J_2 \, (=s)$ of the second
space, and again any spin multiplet $(\J_3,i_3)$ from the third factor space.
Disregarding scalar factors, there is at most one way of coupling these
multiplets to a singlet state. The coupling coefficients are known as $3j$
symbols and the desired singlet is
\begin{eqnarray}
\nonumber
\ket{(\J_1,i_1),\J_2,(\J_3,i_3)}:=\\
\sum_{m_1,m_2,m_3}
\tj{\J_1}{\J_2}{\J_3}{m_1}{m_2}{m_3}\ket{(\J_1,i_1),m_1}\otimes
\ket{\J_2,m_2}\otimes\ket{(\J_3,i_3),m_3}.\label{singlet3j}
\end{eqnarray}
Further below we will be using a graphical language for constructions and actual
calculations. For instance, the singlet \refeq{singlet3j} and its dual are
depicted by three straight lines carrying arrows, see Fig.~\ref{graphsarrows}.
The coupling coefficients of the singlets shown in Fig.~\ref{graphsarrows} are
given in Fig.~\ref{graphsnoarrows}. More precisely, the objects shown
Fig.~\ref{graphsarrows} are obtained by multiplying the objects in
Fig.~\ref{graphsnoarrows} by states $\ket{(\J,i),m}$ (or the dual) and summing over
all magnetic quantum numbers $m$.

The singlet $G$ can be written as superposition of these elementary singlets
\be
G=\sum_{(\J_1,i_1),(\J_3,i_3)}A^{\J_1,\J_3}_{i_1,i_3}\ket{(\J_1,i_1),\J_2,(\J_3,i_3)}
\label{Gdurch3j}
\ee
with suitable coefficients $A^{\J_1,\J_3}_{i_1,i_3}$. Note that $\J_2$ has been
suppressed as index-like argument of $A$ as $\J_2$ is always identical to
$\12$ and unique (for this reason no $i_2$ has been introduced above).
Due to the symmetry of $3j$ symbols with respect to exchange of two columns
\be
\tj{\J_3}{\J_2}{\J_1}{m_3}{m_2}{m_1}=(-1)^{\J_1+\J_2+\J_3}
\tj{\J_1}{\J_2}{\J_3}{m_1}{m_2}{m_3}\label{transpose3j}
\ee
we conclude that
\be
A^{\J_3,\J_1}_{i_3,i_1}=(-1)^{\J_1+\J_2+\J_3}A^{\J_1,\J_3}_{i_1,i_3}.\label{transposeA}
\ee
is a sufficient condition for parity invariance. Note that $\J_1+\J_2+\J_3$ 
is always integer.

Also note that only few combinations ${\J_1}, {\J_2}, {\J_3}$ need to be
considered: if the triangle condition $|\J_1-\J_2|\le \J_3\le|\J_1+\J_2|$ or any
condition obtained by permutations of the indices is violated, the three
multiplets can not couple to a singlet. Since $\J_2=\12$, we are left with the
combinations $\{\J_1, \J_3\}=\{0,\12\}, \{\12,1\}, \{1,\32\},...$. This is a
natural result, since $\J_1$ and $\J_2=\12$ only couple to $\J_1\pm
\12$. Let us denote by $n_\J$ the
number of spin-$\J$ multiplets.
By use of the symmetry \refeq{transposeA} we may reduce all
possible coefficients $A$ to a set of $n_\J \times n_{\J+\s12}$ matrices
$A^{\J,\J+\s12}$ with matrix elements
\be
\Big(A^{\J,\J+\s12}\Big)_{\il,\ir}:=A^{\J,\J+\s12}_{\il,\ir}.
\ee

\subsection{Norm and transfer matrix}

Next, we want to calculate the norm $\bra{\psi}\psi\rangle$ and the
expectation value of the Hamiltonian $\bra{\psi}H\ket{\psi}$ in the
thermodynamic limit. The computation leads to
\be
\langle\psi\ket{\psi}=\Tr (g_1^+g_1\cdot g_2^+g_2\cdot ...\cdot g_L^+g_L),\label{norm}
\ee
where $g^+$ $\in V^*\otimes(\Vs)^*\otimes V$ is the dual of $g$ $\in
V\otimes\Vs\otimes V^*$ and the contraction over the second space is
implicitly understood in $g^+ g$. Hence $T:=g^+ g$ is a linear map 
$V\otimes V^*\to V\otimes V^*$.
For the computation of the norm we employ the
transfer matrix trick yielding for the r.h.s. of \refeq{norm}
\be
\langle\psi\ket{\psi}=\Tr (\underbrace{T\cdot ...\cdot T}_{L\ \mbox{times}})
=\sum_{\Lambda}\Lambda^L,\label{puretransfermatrix}
\ee
where the sum is over all eigenvalues $\Lambda$ of $T$. Obviously, in the 
thermodynamic limit only the largest eigenvalue(s) contribute.

The computation of the leading eigenvalue is facilitated by the singlet nature
of the leading eigenstate. There are not many independent singlet states in 
$V\otimes V^*$. A $(\J,i)$ multiplet in $V$ and a $(\tilde\J,\tilde i)$
multiplet in $V^*$ couple to a singlet iff $\J=\tilde\J$ (with arbitrary $i$
and $\tilde i$). The (normalized) singlet is given by
\be
\sigma(\J;i,\tilde i):=\frac1{\sqrt{2\J+1}}\sum_{m=-\J}^\J\ket{\J,i,m}\otimes\bra{\J,\tilde i,m}.
\ee
Graphicially, this singlet is depicted by a link carrying an arrow pointing
from the $V^*$ to the $V$ space.

\begin{figure}
\begin{center}
\includegraphics[width=0.65\linewidth]{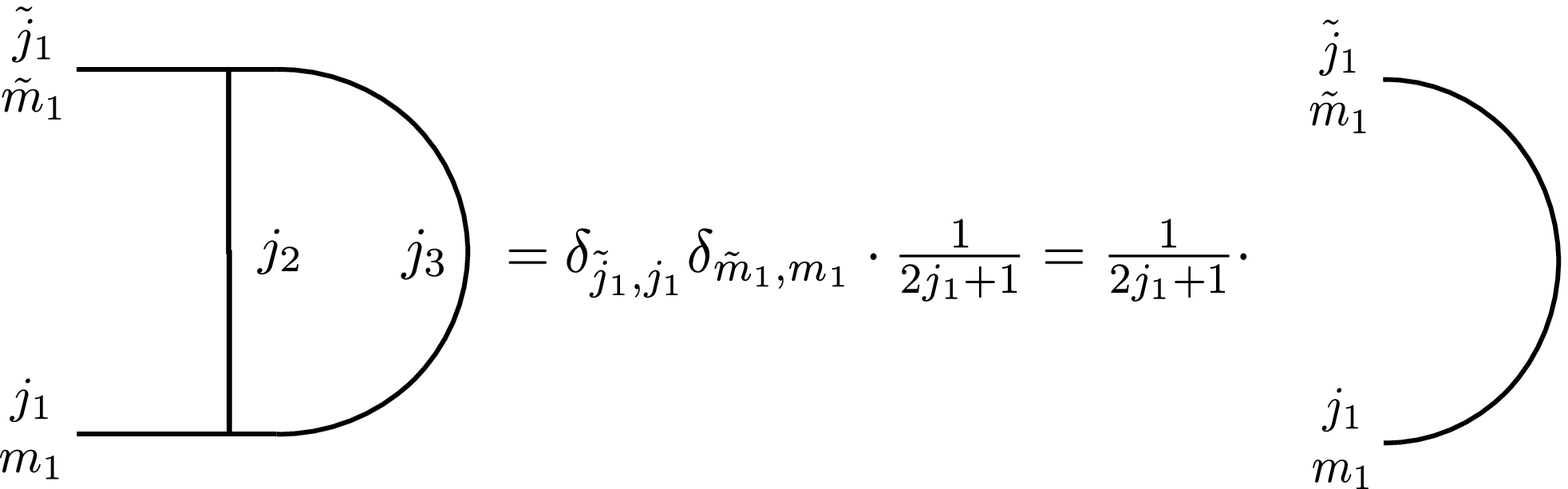}
\end{center}
\caption{Illustration of identity \refeq{3junitarity}: The summation over two
  internal magnetic quantum numbers leads to an ``arc''-singlet.}
\label{s0}
\end{figure}

\begin{figure}
\begin{center}
\includegraphics[width=0.6\linewidth]{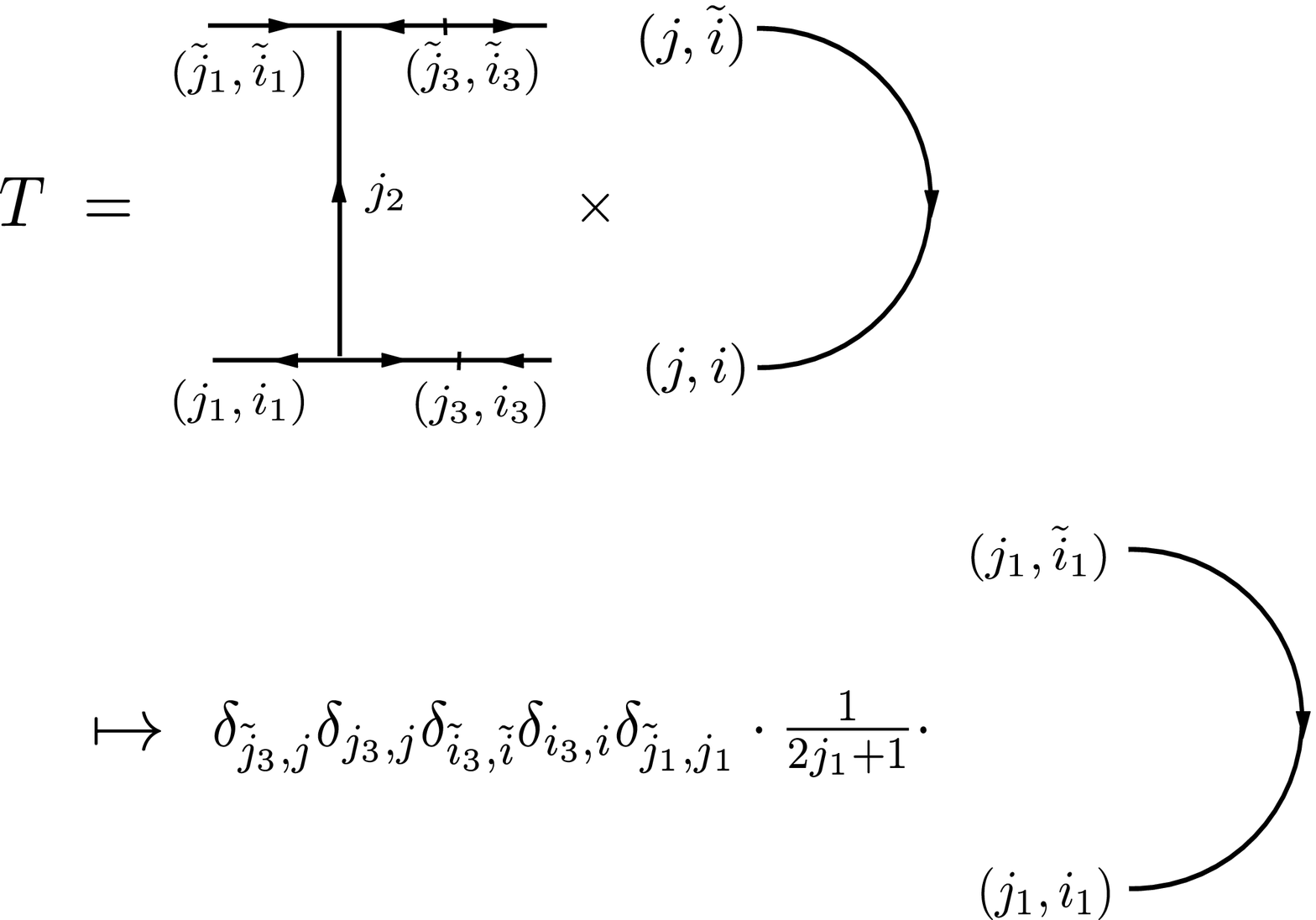}
\end{center}
\caption{Depiction of the transfer matrix $T$ in terms of local vertices and the
  action onto singlet states.}
\label{T}
\end{figure}

The action of the transfer matrix $T$ onto a singlet $\sigma(\J;i,\tilde i)$
produces similar singlets where $\J$ is changed by $\pm \12$. The
$\sigma(\Jl;\il,\til)-\sigma(\Jr;\ir,\tir)$ matrix element of $T$ is
\be
\bra{\Jl;\il,\til}T\ket{\Jr;\ir,\tir}=
\frac1{\sqrt{(2\Jl+1)(2\Jr+1)}}
A^{\Jl,\Jr}_{\il,\ir} \Big(A^{\Jl,\Jr}_{\til,\tir}\Big)^*,\label{Telements}
\ee
where $\Jl=\Jr\pm \12$. (Expression \refeq{Telements} is the analogue of eq.~(6) in
\cite{Dukelsky98JPA}, but here we do not impose condition (10) of 
\cite{Dukelsky98JPA}.) The matrix $T$ has a simple block structure
with zero diagonal blocks and non-zero secondary diagonal blocks.
From \refeq{transposeA} we conclude that the matrix is symmetric. The defining
blocks are
\be
\bra{\J;\il,\til}T\ket{\J+\s12;\ir,\tir}=
\frac1{\sqrt{(2\J+1)(2\J+2)}}
A^{\J,\J+\s12}_{\il,\ir} \Big({A^{\J,\J+\s12}_{\til,\tir}}\Big)^*.\label{Telements2}
\ee
In expression \refeq{Telements} the coefficients $A$ and the
complex conjugate $A^*$ derive from the explicit appearance in
\refeq{Gdurch3j} and the prefactor is obtained from the identity of $3j$
symbols (if $\J_1$ is found in the product of $\J_2$ and $\J_3$)
\be
\sum_{m_2,m_3}
\tj{\J_1'}{\J_2}{\J_3}{m_1'}{m_2}{m_3}\tj{\J_1}{\J_2}{\J_3}{m_1}{m_2}{m_3}
=\frac1{2\J_1+1}\delta_{\J_1,\J_1'}\delta_{m_1,m_1'},\label{3junitarity}
\ee
illustrated in Fig.~\ref{s0}. 
The computation of the matrix elements of $T$ is graphically presented in
Fig.~\ref{T}.

For a given space $V$ with a certain number of spin-0, $\12$, 1, $\32$
... multiplets and a certain set of parameters $A(\Jl,\il;\Jr,\ir)$, the
transfer matrix has to be diagonalized in the singlet space. The total number
of non-zero coefficients $A(\Jl,\il;\Jr,\ir)$ is $\sum_\J n_\J n_{\J+\s12}$
(with sum over $\J=0, \12, 1, \32,...$ and $n_\J$ denoting the number of
spin-$\J$ multiplets). The total dimension of $V\otimes V^*$ is $\big(\sum_\J
n_\J(2\J+1)\big)^2$, but the singlet subspace is much lower dimensional:
$\sum_\J n_\J^2$.  Due to the still high dimensionality, the diagonalisation
in the singlet space has to be done numerically. The eigenvalues come in pairs
$\pm\Lambda$. The two largest eigenvalues and the corresponding eigenstates
determine the physics in the thermodynamic limit.

\subsection{Nearest-neighbour couplings}

We are interested in the spin-$\12$ Heisenberg chain with nearest-neighbour
interaction with Hamiltonian
\be
H=\sum_{l=1}^L \vec S_l \vec S_{l+1}.
\ee
The local Hamiltonian is $h_{l,l+1}=\vec S_l \vec S_{l+1}=1/4-P^{nn}_0$ 
where $P^{nn}_0$ is the
projector onto the nearest-neighbour singlet space. We want to determine the
matrix-product state with minimal expectation value of the total Hamiltonian
$H=\sum_l h_l$. Due to translational invariance this is achieved by
minimizing the expectation value of a single local interaction. In analogy to
\refeq{puretransfermatrix} we obtain 
\be 
\bra{\psi} P^{nn}_0\ket{\psi}=\Tr (\tilde
T\underbrace{T\cdot ...\cdot T}_{L-2\ \mbox{times}})
=\Lambda_0^{L-2}\bra{0}T_2\ket{0},
\label{modiftransfermatrix} 
\ee 
where we assumed $P^{nn}_0$ to act on sites 1 and 2. $T_2$ is a modified
transfer matrix acting in $V\otimes V^*$, $\ket{0}$ is the (normalized)
leading eigenstate
of the transfer matrix $T$ and we kept the only term dominating in the
thermodynamical limit. Hence
\be
\frac{\bra{\psi} P^{nn}_0\ket{\psi}}{\langle\psi\ket{\psi}}
=\Lambda_0^{-2}\bra{0}T_2\ket{0}.
\ee
\begin{figure}
\begin{center}
\includegraphics[width=0.35\linewidth]{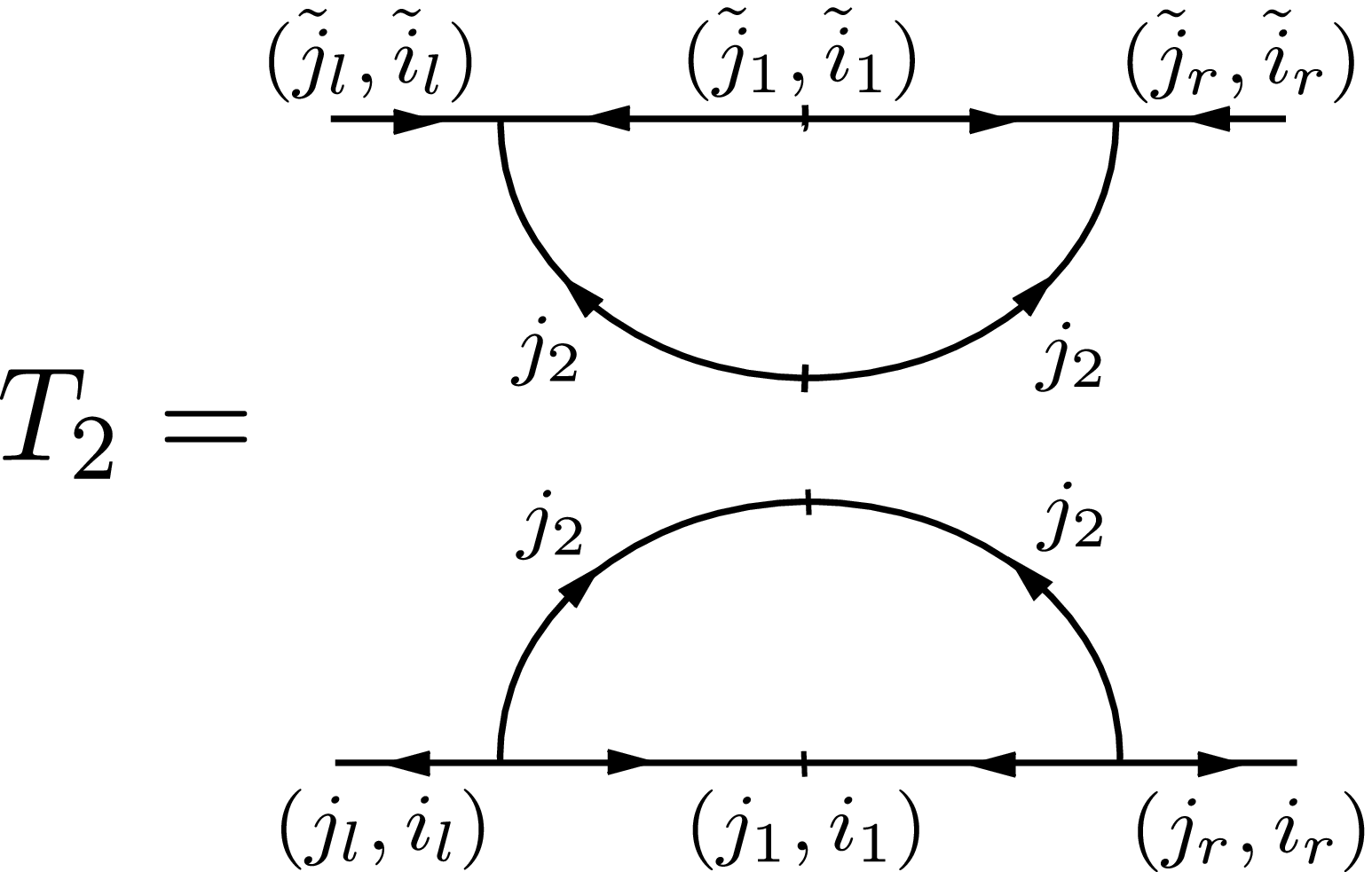}
\end{center}
\caption{Depiction of the modified matrix $T_2$ in terms of local vertices.}
\label{T2}
\end{figure}
\begin{figure}
\begin{center}
\includegraphics[width=0.65\linewidth]{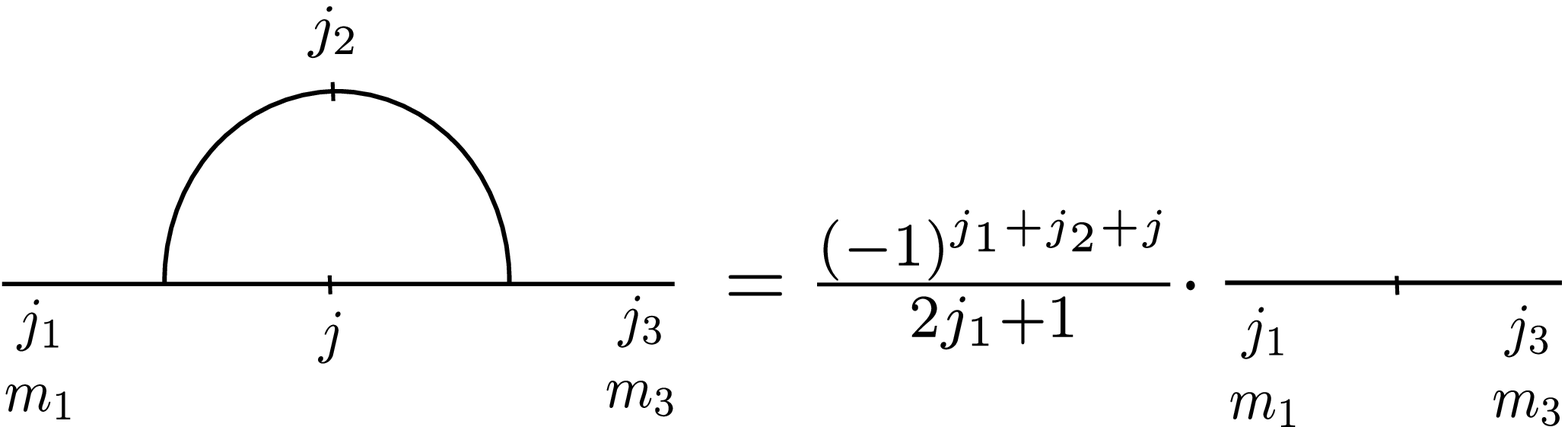}
\end{center}
\caption{The summation over two internal magnetic quantum numbers leads to
an ``edge''-singlet.}
\label{singlet}
\end{figure}
The computation of the matrix elements of $T_2$ is
described graphically in Figs.~\ref{T2} and \ref{singlet}. 
In contrast to the transfer matrix $T$ the
modified matrix $T_2$ is block diagonal with
$\sigma(\J;\il,\til)-\sigma(\J;\ir,\tir)$ matrix element 
\be
\bra{\J;\il,\til}T_2\ket{\J;\ir,\tir}=\frac 1{(2\J+1)^2}
B^{\J,\J}_{\il,\ir} \Big(B^{\J,\J}_{\til,\tir}\Big)^*.\label{T2elements}
\ee
The matrices $B$ are given by
\be
B^{\J,\J}
=\sum_{\J_1}(-1)^{\s12+\J+\J_1}\label{Belements}
A^{\J,\J_1}A^{\J_1,\J},
\ee
where only the values $\J_1=\J\pm \12$ lead to non-zero
terms. (Equations (\ref{T2elements},\ref{Belements})
are the analogue of eq.~(26) in \cite{Dukelsky98JPA}.) Using this and
the (sufficient) condition \refeq{transposeA} for parity invariance we find
\be
B^{\J,\J}=\Big[A^{\J-\s12,\J}\Big]^T A^{\J-\s12,\J}
+A^{\J,\J+\s12}\Big[A^{\J,\J+\s12}\Big]^T.
\ee

\subsection{Next-Nearest-neighbour couplings}

The next-nearest neighbour interactions are manageable, too.  In the
thermodynamical limit we find
\be 
\frac{\bra{\psi}
  P^{nnn}_0\ket{\psi}}{\langle\psi\ket{\psi}} =\Lambda_0^{-3}\bra{0}T_3\ket{0}.
\ee 
\begin{figure}
\begin{center}
\includegraphics[width=0.4\linewidth]{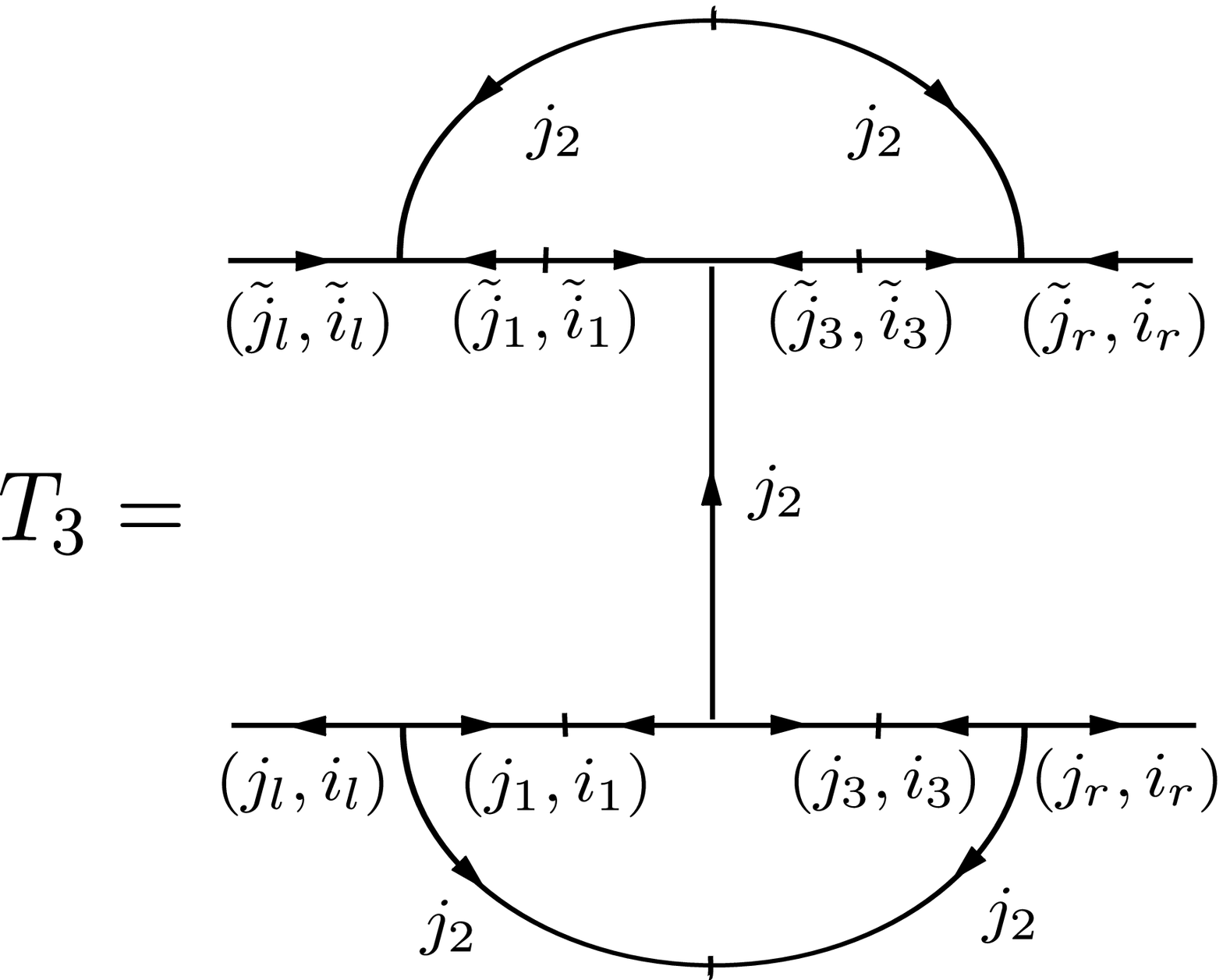}
\end{center}
\caption{Depiction of the modified matrix $T_3$ in terms of local vertices.}
\label{T3}
\end{figure}
\begin{figure}
\begin{center}
\includegraphics[width=0.8\linewidth]{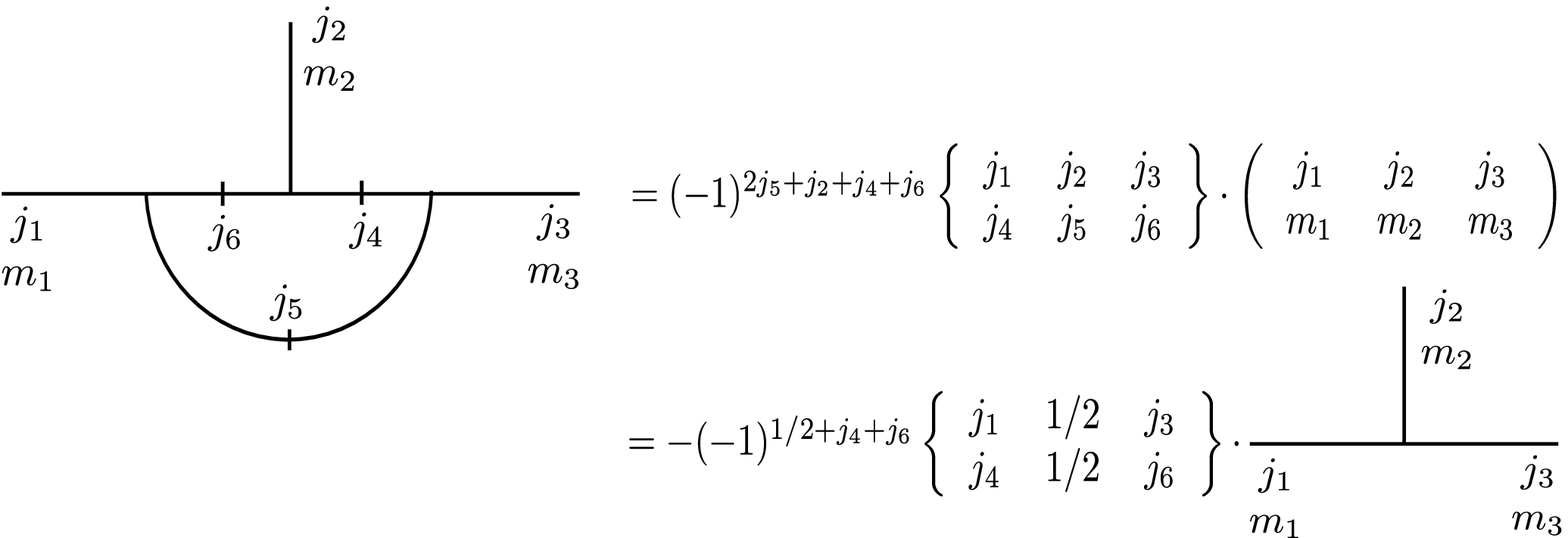}
\end{center}
\caption{The summation over three internal magnetic quantum numbers leads to
a $6j$ symbol times an elementary vertex.}
\label{6j}
\end{figure}
The computation of the matrix elements of $T_3$ is described graphically in
Fig.~\ref{T3} and Fig.~\ref{6j}. 
In contrast to the modified transfer matrix $T_2$, but like $T$, the
matrix $T_3$ has zero diagonal blocks and non-zero secondary diagonal blocks.
The $\sigma(\Jl;\il,\til)-\sigma(\Jr;\ir,\tir)$ matrix element of $T_3$ is
\be
\bra{\Jl;\il,\til}T_3\ket{\Jr;\ir,\tir}=
\frac1{\sqrt{(2\Jl+1)(2\Jr+1)}}
C^{\J,\J+\s12}_{\il,\ir} \Big({C^{\J,\J+\s12}_{\til,\tir}}\Big)^*.\label{T3elements}
\ee
where $\Jl=\Jr\pm \12$ and $C^{\Jl,\Jr}$ is given by
\be
C^{\Jl,\Jr}=\sum_{\J_1,\J_3}(-1)^{\s12+\J_1+\J_3}\sj{\Jl}{\s12}{\Jr}{\J_3}{\s12}{\J_1}
A^{\Jl,\J_1}A^{\J_1,\J_3}A^{\J_3,\Jr}.\label{Celements}
\ee
(There is no analogue to eqs.~(\ref{T3elements},\ref{Celements}) in
\cite{Dukelsky98EPL,Dukelsky98JPA}.)
Only three combinations of $\J_1,\J_3$ yield non-vanishing contributions. 
For $\Jr=\Jl+1/2$
only $(\J_1,\J_3)=$ $(\Jl-1/2,\Jl)$, $(\Jl+1/2,\Jl)$, $(\Jl+1/2,\Jl+1)$ are
relevant. From this and
the (sufficient) condition \refeq{transposeA} for parity invariance we find
\begin{eqnarray}
\nonumber
C^{\J,\J+\s12}&=\sj{\J}{\s12}{\J+\s12}{\J}{\s12}{\J-\s12}
\Big(A^{\J-\s12,\J}\Big)^TA^{\J-\s12,\J}A^{\J,\J+\s12}\\\nonumber
&+
\sj{\J}{\s12}{\J+\s12}{\J}{\s12}{\J+\s12}
A^{\J,\J+\s12}\Big(A^{\J,\J+\s12}\Big)^TA^{\J,\J+\s12}\\
&+
\sj{\J}{\s12}{\J+\s12}{\J+1}{\s12}{\J+\s12}
A^{\J,\J+\s12}A^{\J+\s12,\J+1}\Big(A^{\J+\s12,\J+1}\Big)^T,
\end{eqnarray}
where the $6j$ symbols evaluate to
\begin{eqnarray}
\nonumber
\sj{\J}{\s12}{\J+\s12}{\J}{\s12}{\J-\s12}=\frac{(-1)^{2\J+1}}{2\J+1},\\\nonumber
\sj{\J}{\s12}{\J+\s12}{\J}{\s12}{\J+\s12}=\frac{(-1)^{2\J+1}}{(2\J+1)(2\J+2)},\\
\sj{\J}{\s12}{\J+\s12}{\J+1}{\s12}{\J+\s12}=\frac{(-1)^{2\J}}{2\J+2},
\end{eqnarray}
with an exception for $\J=0$ where the first listed 
$6j$ symbol has to be taken as 0.

\section{Results}\label{Results}

For the nearest-neighbour spin-$\12$ Heisenberg chain we found that already a
few low-dimensional multiplets in the matrix space $V$ yield excellent
results, e.g.~the ground-state energy differs from the exact result
$e_0=1/4-\ln 2$ by about $5\cdot 10^{-5}$ and dimer correlations are off the
exact results by about $10^{-4}$. This is achieved with $n_0=4$,
$n_{\s12}=4$, $n_{1}=3$, $n_{\s32}=2$, $n_{2}=1$ ($n_s$=0 for $s>2$). Note
that the attempt to include higher spin multiplets at the expense of reducing
the low-spin multiplets is not successful as for instance $n_s=1$ for all $s$
leads to the simple dimer (Majumdar-Ghosh) state.

The minimization of the energy expectation value yields the following list of
coefficients
\begin{eqnarray}
\nonumber
A^{0,\s12}&=\left[ \begin {array}{cccc}  1.0&0&0&0\\ \noalign{\medskip}0&-
 0.283412&0&0\\ \noalign{\medskip}0&0& 0.183208&0
\\ \noalign{\medskip}0&0&0& 0.679798\end {array} \right],\quad\\\nonumber
 A^{\s12,1}&=
\left[ \begin {array}{ccc}  0.737765&0&0\\ \noalign{\medskip}
 0.257549&- 0.860519&0\\ \noalign{\medskip} 0.610288&-
 0.093768& 0.050733\\ \noalign{\medskip} 0.191246&-
 0.631258& 0.181757\end {array} \right]\\
A^{1,\s32}&=
\left[ \begin {array}{cc}  0.510676&0\\ \noalign{\medskip}-
 0.5075510& 0.714082\\ \noalign{\medskip}- 0.292858&
 0.712182\end {array} \right],
\qquad
A^{\s32,2}=
\left[ \begin {array}{c}  0.265736\\ \noalign{\medskip}-
 0.607939\end {array} \right] 
\label{ExplCoeff0}
\end{eqnarray}
and a value of the ground-state energy which
compares well with the exact value \cite{Hul38}
\be
e_0^{\rm num}=-0.443\ 092\ 175...,\qquad e_0^{\rm exact}=-0.443\ 147\ 180...\label{comparee0}
\ee
Note that
$A^{0,\s12}$ is strictly diagonal, and the other matrices have strictly zero
entries above the diagonal due to a ``gauge'' freedom. The MPS is invariant
under a transformation $A^{\Jl,\Jr}\to{\cal{O}}_{\Jl}^{-1}A^{\Jl,\Jr}{\cal{O}}_{\Jr}$
where ${\cal{O}}_\J$ are arbitrary orthogonal $n_\J\times n_\J$-matrices. Also
note that the entries of the matrices are strictly real.

\begin{figure}
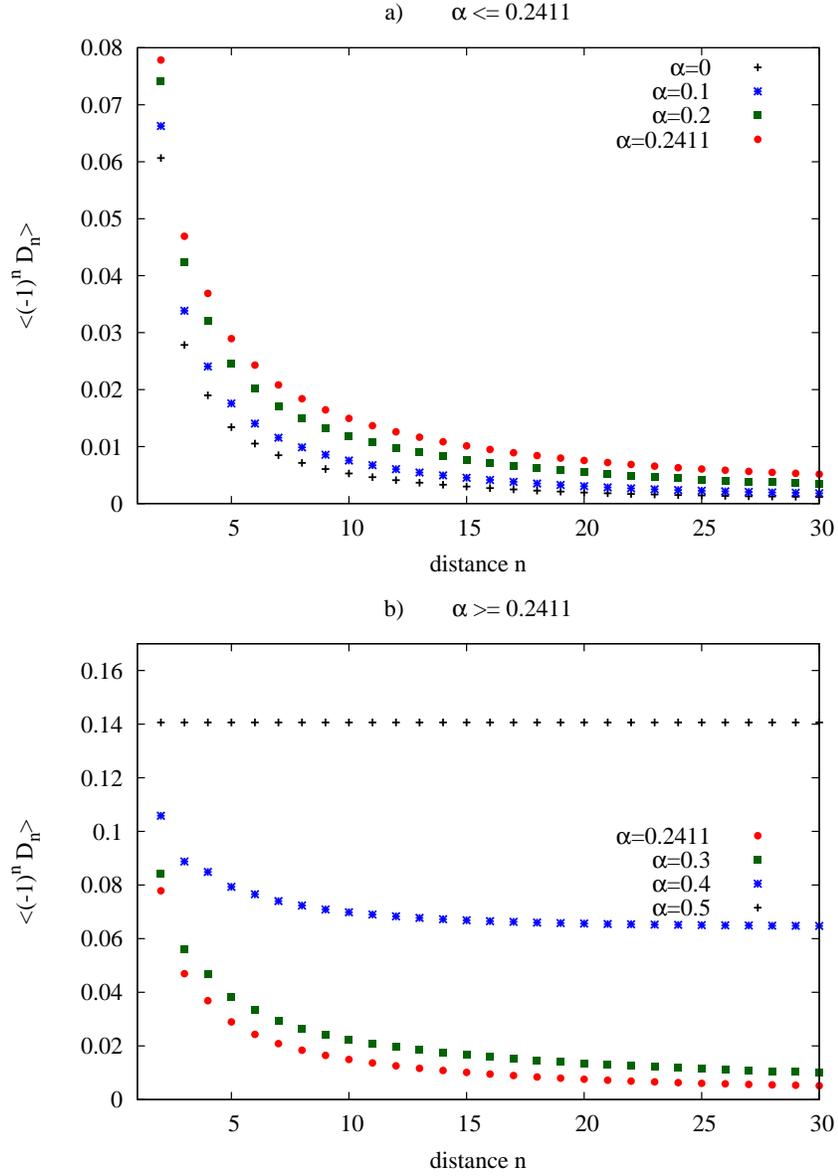

\begin{center}
\includegraphics[width=0.7\linewidth]{dimer_all.epsi}\vskip0.2cm
\includegraphics[width=0.7\linewidth]{dimer_all2.epsi}
\end{center}
\caption{Dimer-dimer correlations of the spin-$\12$ Heisenberg chain with
  frustration parameter $\alpha$: Linear plots of $(-1)^n D_n$ for a)
  $\alpha=0, 0.1, 0.2, 0.2411$ and b) $\alpha=0.2411, 0.3, 0.4, 0.5$. Note the
  long-range order in plot b) for $\alpha>0.2411$.}
\label{graphdimeral}
\end{figure}

Next we give numerical results for the dimer-dimer correlation function
\be
\tilde D_n:=\langle(\vec S_1\cdot\vec S_2)(\vec S_{n+1}\cdot\vec S_{n+2})\rangle.
\ee
For $n\to\infty$ this converges to $e_0^2$, hence it is more instructive to
study the connected dimer correlations
\be
D_n:=\tilde D_n-\tilde D_\infty.
\ee
For $n=$ 2, 3, 4 the values are known exactly 
\cite{Sato06}
which we use for
comparison with our MPS calculations
\begin{eqnarray}
\nonumber
D_2^{\rm num}&= +0.060\ 639...,\qquad & D_2^{\rm exact}= +0.060\ 824...,\\
D_3^{\rm num}&= -0.027\ 838..., & D_3^{\rm exact}= -0.027\ 737...,\\\nonumber
D_4^{\rm num}&= +0.018\ 986..., & D_4^{\rm exact}= +0.018\ 928... .
\end{eqnarray}
We expect that the absolute numerical accuracy is similar also for larger
distances $n$ of the local dimer operators. These results are plotted in
Fig.~\ref{graphdimeral} as $(-1)^n D_n$ versus $n$. Note that all $(-1)^n D_n$
are positive which implies a sublattice structure with sign alternation of
the correlations. There is no
long-range order for the spin-$\12$ Heisenberg chain with nearest-neighbour
interactions.

\begin{figure}
\begin{center}
\includegraphics[width=0.7\linewidth]{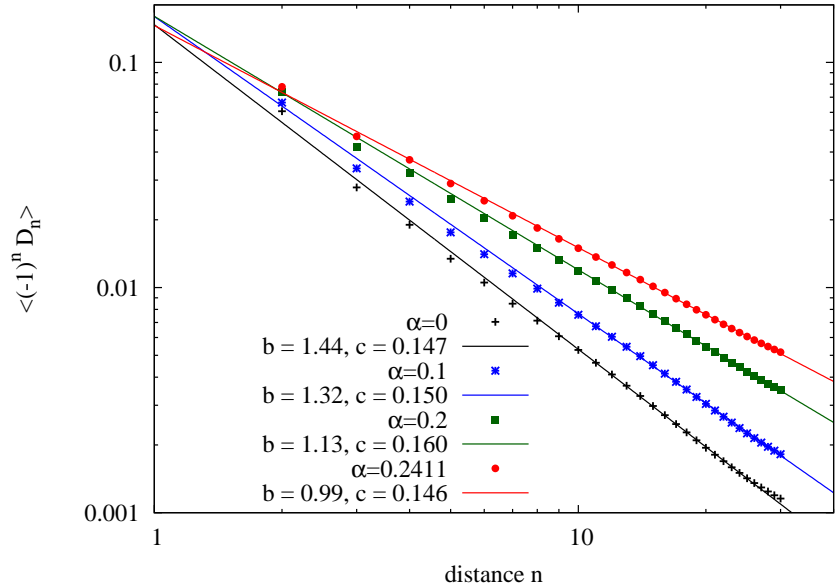}
\end{center}
\caption{Dimer-dimer correlations of the spin-$\12$ Heisenberg chain with
  frustration parameter $\alpha\le 0.2411$: Double log plot of $(-1)^n D_n$
  versus distance $n$ and algebraic lines $c/n^b$ from fits to the data for
  $n$ in the range $[10,25]$.}
\label{graphdimer}
\end{figure}

Next we are interested in the frustrated spin-$\12$ Heisenberg chain with
Hamiltonian
\be
H=\sum_{l=1}^L \left(\vec S_l \vec S_{l+1}+\alpha\vec S_l \vec S_{l+2}\right).
\ee
The system shows algebraically decaying dimer-dimer correlations for
$\alpha\le 0.2411...$, see Fig.~\ref{graphdimer}, and long-range dimer order 
for $\alpha>0.2411...$, see Fig.~\ref{graphdimeral} b). 
The critical value $\alpha_c=0.2411$ was established 
in \cite{OkamotoNomura92,Eggert96}. The dimer-dimer
correlations are fitted well by algebraic curves for all $\alpha\le 0.2411..$
with $\alpha$-dependent exponent. From field theoretical considerations the
exponent is expected to be identical to 1. We attribute the deviations to
logarithmic corrections for $\alpha< 0.2411..$ which apparently vanish at
$\alpha_c$.

As an illustration of spin-spin-correlation functions we present results for
the nearest- and next-nearest-neighbour cases in table \ref{my_table}. 
For $\alpha=0$ the exact values are known \cite{Hul38} $C_2^{\rm
  exact}=e_0^{\rm exact}$, see
\refeq{comparee0}, and \cite{Sato06} $C_3^{\rm exact}=0.182\ 039...$. Even for $C_3$ our
numerical value deviates from the exact value less than $10^{-4}$.
Due to the variational nature of our calculations the expectation values of the
energy are strict upper bounds for the ground-state energy. Note that for cases
$\alpha=0.1, 0.2411, 0.3, 0.4$ our results are lower than those given in
\cite{Chitra95} with small differences ranging from $10^{-4}$ to
$10^{-3}$. Hence, the deviation of these DMRG results from the true ground-state
energy must be of the same order or even larger. In DMRG calculations there
are two sources of errors: (i) truncation of the Hilbert space and (ii) finite-size
effects due to the finite length of the considered chains. In our approach we
deal with the strictly infinitely long chain.
\begin{table}
\begin{center}
\item[]\begin{tabular}{c|ccccccc}
$\alpha$ & 0.0 & 0.1 & 0.2 & $\alpha_c$ & 0.3 & 0.4 & 0.5  \\
\hline
$C_2$ & -0.443092 & -0.442655 & -0.440916 & -0.439574 &-0.436475&
-0.420659 & -0.375000\\
$C_3$ & 0.181942 & 0.173570 & 0.162233 & 0.156176 & 0.144794 & 0.100870 &
0.0 \\
$e_0$ & -0.443092 & -0.425298 & -0.408469 & -0.401920 &-0.393037 &
-0.380311 & -0.375000 
\end{tabular}
\end{center}
%
\caption{Numerical values for the correlations $C_n:=\langle \vec S_1 \vec
  S_{n+1}\rangle$ for $n=1, 2$ and various values of the frustration
  parameter $\alpha$. The ground-state energy $e_0=C_1+\alpha C_2$ is given in
  the last row.}
\label{my_table}
\end{table}
The numerical computations of the data presented in this section were done by
use of Maple 13 on a laptop computer. The total computation for the seven
cases of the frustration parameter took about 1 hour.

A more complete study of the correlation functions and of the physics of frustrated
systems with $\alpha>0.5$ will be presented elsewhere. Here we like to note that
for $\alpha>0.5$ the matrices replacing \refeq{ExplCoeff0} will contain
intrinsically complex numbers.

\section{Conclusion}

We showed how to employ systematically $su(2)$ invariance for matrix product
states and how to carry out the variational computation of the ground state
energy in a numerically most efficient manner. The algebraic computations for
the $su(2)$-invariant bulding blocks were put in diagrammatic formulation.  As
an example we used the (frustrated) spin-$\12$ Heisenberg chain with nearest-
and next-nearest neighbour interaction. Our algebraic constructions led to the
main results \refeq{Telements} for the transfer matrix, and
(\ref{T2elements},\ref{Belements}) and (\ref{T3elements},\ref{Celements}) for
the modified transfer matrices, where the coefficient matrix has to satisfy
the relation \refeq{transposeA} for parity invariance.

Our calculations are very similar to those of
\cite{Dukelsky98EPL,Dukelsky98JPA} who applied the method
to gapped spin-1 chains and spin
ladders. The variational MPS calculations for the spin-$\12$ Heisenberg chain
are demanding on their own: the model shows algebraically decaying
correlation functions and it is by no means clear if signatures of this decay
can already be seen in variational MPS calculations with only few multiplets
in the matrix space. Also, in contrast to
\cite{Dukelsky98EPL,Dukelsky98JPA}, the matrix space we had to deal with 
consists of all integer and half-odd integer spin multiplets.
In our concrete calculation we used a matrix space composed 
of 4 singlets, 4 doublets, 3 triplets, 2
quadruplets, and 1 quintuplets. We managed to calculate the ground-state energy
within a precision better than $10^{-4}$. Also, the correlation functions were
computed within an accuracy of the order $10^{-4}$ and allowed for the
identification of the scaling dimension in the case of critical frustration
($\alpha=0.2411$). 

The actual numerical
calculations like matrix diagonalizations were reduced by the algebraic
$su(2)$ implementation from a 1156-dimensional to a 46-dimensional space. The
systematic inclusion of more and higher spin multiplets is obvious.
Generalizations of these calculations are straight forward,~e.g. to spin-$S$
Heisenberg chains with competing interactions.  We are convinced that a
systematic application of symmetries to the MPS analysis of quantum spin chains
will provide high quality data with only small truncation errors.

\subsection*{Acknowledgments.}  The authors would like to thank F. G\"ohmann,
M. Karbach and J. Sirker for valuable discussions, and J. Dukelsky and
A. Schadschneider for relevant information on related work.  The authors
gratefully acknowledge support by {\em Deutsche Forschungsgemeinschaft} under
project {\em Renormierungsgruppe KL 645/6-1}.

\end{document}